\documentclass{article}
\usepackage{graphicx} 
\usepackage{fancyhdr}
\usepackage{booktabs}
\usepackage{geometry}
\usepackage{amsthm}
\usepackage{amsmath}
\usepackage{amssymb}
\usepackage{hyperref}
\usepackage{stmaryrd}
\usepackage{dsfont}
\usepackage{tabularx}
\usepackage{pdfpages}
\usepackage[round]{natbib}
\usepackage{authblk}
\usepackage{mathtools}
\usepackage{subcaption}

\newcommand{\R}{\mathbb{R}}
\newcommand{\thetab}{{\pmb \theta}}
\newcommand{\ub}{\ensuremath{\mathbf{u}}}
\newcommand{\n}{^{(n)}}
\newcommand{\xb}{\ensuremath{\mathbf{x}}}

\newcommand{\Xb}{\mathbf{X}}
\newcommand{\Yb}{\mathbf{Y}}
\newcommand{\Zb}{\mathbf{Z}}

\newcommand{\thetat}{{\widehat\thetab}^{\phantom n}{\phantom\theta}\hspace{-3.0mm}^{{(n)}}}

\theoremstyle{definition}

\usepackage{stackengine}
\stackMath
\newcommand\tenq[2][1]{%
\def\useanchorwidth{T}%
\ifnum#1>1%
\stackunder[0pt]{\tenq[\numexpr#1-1\relax]{#2}}{\scriptstyle\thicksim}%
\else%
\stackunder[1pt]{#2}{\scriptscriptstyle\thicksim}%
\fi%
}

\title{Discussion of  ``Robust Distance Covariance" \\ by
S.~Leyder, J.~Raymaekers,   and P.J.~Rousseeuw}

\author[1,2]{Hallin Marc}
\author[3]{Davide La Vecchia}
\author[4]{Hang Liu }
\author[4]{Xinyi Xu}

\affil[1]{Department of Mathematics, Universit\'e Libre de Bruxelles, Belgium}

\affil[2]{Institute of Information Theory and Automation, Czech Academy of Sciences, Prague, Czech Republic}
\affil[3]{Geneva School of Economics and Management, University of Geneva}
\affil[4]{Department of Statistics and Finance, School of Management, University of Science and Technology of China}

\date{}

\numberwithin{equation}{section}

\theoremstyle{definition}
\newtheorem{definition}{Definition}[section]

\theoremstyle{remark}

\DeclareMathOperator*{\argmin}{argmin}

\begin{document}

\maketitle

\section{Introduction}

\subsection{Distance covariance and the null hypothesis of independence}

Testing the null hypothesis ${\cal H}_0$ of independence between two variables $X$ and $Y$ is an important and fundamental problem in theoretical as well as in applied statistics. Thanks to the fact that they vanish if and only if   $X$ and $Y$ are independent, the {\it distance covariance} dCov$(X,Y)$ and the corresponding {\it distance correlation} dCor$(X,Y)$ are natural measures of the dependence   between~$X$ and~$Y$. Their empirical versions  constitute tests statistics for testing~${\cal H}_0$ against the alternative~${\cal H}_1$ of unspecified forms of dependence. Introduced by \cite{szekely2007measuring}, distance covariance and correlation have been thoroughly studied, 
and are  used in a number of applications. In the sequel, we concentrate on distance covariance, but distance correlation clearly enjoys similar properties.

A remark about the notation is in order here. The notation dCov$(X,Y)$ is somewhat misleading inasmuch as dCov is a function of the joint distribution ${\rm P}_{\!(X,Y)}$ of $X$ and $Y$,   not  a measurable function of $X$ and $Y$. The empirical version of dCov$(X,Y)$, based on a sample $$\big({\bf X}\n, {\bf Y}\n\big)\coloneqq \big((X\n_1,Y\n_1),\ldots, (X\n_n,Y\n_n)\big),$$ 
will be denoted as dCov$({X}\n, {Y}\n)$ and actually is the function dCov computed at the empirical distribution of the random $n$-tuple $(X\n_1,Y\n_1),\ldots, (X\n_n,Y\n_n)$.

Despite the popularity of the concept, the robustness properties of the empirical distance covariance dCov$({\bf X}\n, {\bf Y}\n)$ remain largely unexplored, though. The present paper by Leyder,  Raymaekers,   and Rousseeuw---below referred to as [LRR]---is a welcome addition to the literature, palliating this  important gap  while revealing some intriguing features and proposing ingenious solutions. Three  results are of particular interest: 
\begin{enumerate}
\item[(i)] an explicit form (Proposition~1 and Corollary~2) of the influence functions of dCov$({X}\n, {Y}\n)$ and~dCor$({X}\n, {Y}\n)$, which are bounded but not redescending (Propositions~2 and~4);
\item[(ii)] an example (Section~3) of a statistic with bounded influence function yet  a breakdown value~$1/n$ tending to zero as $n\to\infty$;
\item[(iii)] a clever trick (the {\it biloop transformation}) of transforming the data in order to obtain a bounded and redescending influence function.
\end{enumerate} 

\subsection{Robustness by transformation}
Point (iii) above highlights what the authors call ``robustness by transformation,''  that is, the fact that, for any bijective transformations $\psi_1$ and $\psi_2$, $X$ and~$Y$ are independent if and only if~$\psi_1(X)$ and $\psi_2(Y)$ are, so that dCov$(X,Y) = 0$  if and only if~dCov$(\psi_1(X),\psi_2(Y)) = 0$. This is an indication that a non-robust characterization of independence  can be turned into a robust one via adequate transformations, and a simple but quite valuable finding of the authors is that these transformations~$\psi_1$ and $\psi_2$ need not be from $\mathbb R$ to $\mathbb R$.  

Transformations, moreover, also can be used to avoid the moment assumptions  required when consi\-dering~dCov$(X,Y)$. This is achieved, for instance, by  the  classical probability integral transformations $\psi_1 = F_X$ and~$\psi_2 = F_Y$ (where $F_X$ and $F_Y$ stand for the distribution functions of $X$ and  $Y$, respectively), the empirical versions $F\n_X$ and $F\n_Y$ of which yield the ranks of the observations $X_1,\ldots,X_n$ and~$Y_1,\ldots,Y_n$. The resulting dCov$(F\n_X(X), F\n_Y(Y))$ then achieves, under the null hypothesis of independence,  distribution-freeness (for absolutely continuous $X$ and $Y$). That transformation can be combined with the use of scores $J_1$ and $J_2$ mapping [0,1) to the real line, yielding  rank-based test statistics   dCov$(J_1\circ F\n_X(X), J_2\circ F\n_Y(Y))$. The case of Gaussian scores $J_1=\Phi^{-1} = J_2$ (where $\Phi$, as usual, denotes the standard normal distribution function), which is briefly  mentioned by [LRR] in their Section~4, corresponds to the traditional (van der Waerden or) Gaussian-score rank statistics; but other scores are possible, such as the Wilcoxon scores~($J_1 (u) = u = J_2(u)$) or the sign test scores~($J_1 (u)=\text{sign}(u-1/2)=J_2(u)$).

Now, as usual, there is a trade-off between robustness and efficiency: depending on the choice of the~$\psi_1$ and $\psi_2$ transformations and    the scores $J_1$ and $J_2$, what are, in terms of efficiency, the costs\linebreak (the benefits) of transformation-robustifying the ``plain version''   dCov$(X\n,Y\n) $ of the empirical distance co\-variance 
 into~dCov$(\psi_1(X\n),\psi_2(Y\n))$ or~dCov$(J_1\circ~\!F\n_X(X)), J_2\circ F\n_Y(Y))$? The answer is not immediate, and that issue, in [LRR], is not considered. Some power features of the rank-based transformations, such as Chernoff-Savage or Hodges-Lehmann results, could be obtained along the lines developed in \cite{HPd2008} for the traditional Wilks test of independence. 

  
\subsection{Beyond bivariate independence}
While the introduction, by  \cite{szekely2007measuring}, of  distance covariance was aiming at characterizing independence between vectors (i.e., $({\bf X},{\bf Y})$ taking values in ${\mathbb R}^{d_1}\times {\mathbb R}^{d_2}$ with $\max(d_1,d_2)>1$), [LRR] are limiting their investigation to  the bivariate case $d_1 =1=d_2$. Their introduction briefly mentions that multivariate ${\bf X}$ and ${\bf Y}$ simply should be replaced with their Euclidean norms~$\Vert {\bf X}\Vert$ and~$\Vert {\bf Y} \Vert$; how to perform that replacement is not completely clear, though. Nor are the applicability and the performance of the [LRR] approach   beyond the case of bivariate $(X,{Y})$. Below, we discuss the possibility of extending distance covariance and   [LRR]'s robustness results to  ${\bf X}$ and ${\bf Y}$ taking values in ${\mathbb R}^d$, on (hyper)spheres, or on more general Riemannian manifolds. 

%
%
%
%
%
%
%

Actually, a rank-based extension of distance covariance for random vectors has been developed by Shi et al.~(2022a and~b)  and \cite{DebSen23}, the results of whom we rapidly mention in Section~\ref{SecRd}.  In Section~\ref{Sec3}, we further extend their multivariate  approach to directional observations---taking values on (hyper)spheres. Extensions to more general compact Riemannian manifolds follow along similar lines, 
and we are currently working on it, but is not without some pitfalls.  For instance, the equivalence between  zero distance covariance  and independence (the fundamental property   justifying distance-covariance-based tests of independence in ${\mathbb R}^d$), in general,  no longer holds for variables taking values on (hyper)spheres: see Section~\ref{Sec:DataSph}. The central idea of our approach is a combination of the distribution-free concepts of multivariate  ranks and signs defined in \cite{hallin2024nonparametric} with distance covariance, via   a suitable data transformation from hyperspheres to tangent spaces---a further illustration of the power of an adequate application of the ``robustness by transformation'' idea developed in [LRR].

%
%
%

\section{Distance covariance  in ${\mathbb R}^d$}\label{SecRd}
Distance covariance in  \cite{szekely2007measuring} was introduced for $({\mathbb R}^{d_1}\times {\mathbb R}^{d_2})$-valued data. Building on the concepts introduced in \cite{Chernoetal17} and \cite{hallin2021center}, the corresponding rank-based versions only came some ten years later, with  Shi et al.~(2022a) and~(2022b) and, in a slightly different form, with \cite{DebSen23}. We only briefly sketch their definitions (Section~\ref{sec21}), and suggest (Section~\ref{sec22}) how to robustify them, along the same lines as~in~[LRR], into statistics with redescending influence functions.\phantom{\cite{ShiEtAl2022a, ShiEtAl2022b}}

\subsection{Rank-based distance covariance in ${\mathbb R}^d$}\label{sec21}
A major difficulty
 stems from the fact that ${\mathbb R}^{d}$, for $d>1$, is not canonically ordered, so that  the very  definition of ranks is problematic. None of the many attempts---marginal ranks, depth-based ranks, spatial ranks, Mahalanobis or elliptical ranks, \ldots we refer to the online appendix of  \cite{hallin2021center} for details and references---to provide a  concept of rank in this multivariate context was fully satisfactory until measure-transportation-based definitions were considered.  \cite{Chernoetal17} and, under the name of  {\it center-outward distribution function},   \cite{hallin2021center} consider   the (a.s.) unique gradient of convex function ${\bf F}_\pm\equiv {\bf F}_{{\bf Z};\pm}$ pushing the distribution~$\rm P$ of some~$d$-dimensional random variable~${\bf Z}$ forward to the spherical uniform ${\rm U}_d$ over the unit ball~${\mathbb S}_d$ and show that it enjoys the typical  properties of a distribution function.  

Based on a sample ${\bf Z}^{(n)}_1,\ldots,{\bf Z}^{(n)}_n$ of i.i.d.\ copies of~${\bf Z}\sim{\rm P}$,  \cite{hallin2021center} also propose  empi\-rical versions   of ${\bf F}_\pm$. For instance, the empirical optimal transport ${\bf F}\n_{0;\pm}$ of the sample  to a grid of~$n$ i.i.d.\ points with spherical uniform distribution ${\rm U}_d$ is a consistent (in the Glivenko-Cantelli sense) estimator of ${\bf F}_\pm$. But, for finite $n$, ${\bf F}\n_{0;\pm}({\bf Z})$ does not enjoy   the independence property of ${\bf U}\sim{\rm U}_d$, namely, independence between~$\Vert{\bf U}\Vert$ and   ${\bf U}/\Vert{\bf U}\Vert$:~$\Vert{\bf F}\n_{0;\pm}({\bf Z})\Vert$ and~${\bf F}\n_{0;\pm}({\bf Z})\n/\Vert{\bf F}_{0;\pm}({\bf Z})\Vert$ are only asymptotically  mutually independent. Also, ${\bf F}\n_{0;\pm}$ does not decompose into mutually independent ranks and signs, and does not provide empirical quantile contours. A more structured grid, which involves a factorization of $n$ into $n=n_Rn_S + n_0$ where $n_R$, $n_S$, and $n_0$ are integers such that~$n_0 < \min(n_R,n_S)$, provides an empirical and still  Glivenko-Cantelli-consistent (as both $n_R$ and~$n_S$ tend to infinity)  version ${\bf F}\n_{\pm}$ of~${\bf F}_\pm$. Contrary to  ${\bf F}\n_{0;\pm}$, ${\bf F}\n_{\pm}$ enjoys, for any~$n$  (and with a tie-breaking procedure in case $n_0 >1$) the independence property of ${\bf U}\sim{\rm U}_d$.  \cite{hallin2021center}  then  define  the rank and the sign  of ${\bf Z}^{(n)}_i$ as~$R^{(n)}_i\coloneqq (n_R +1)\Vert {\bf F}^{(n)}_\pm({\bf Z}^{(n)}_i)\Vert$ and~${\bf S}^{(n)}_i\coloneqq {\bf F}^{(n)}_\pm({\bf Z}^{(n)}_i) / \Vert {\bf F}^{(n)}_\pm({\bf Z}^{(n)}_i)\Vert$, respectively.  We refer to \cite{hallin2021center} for details and to \cite{hallin2022measure} for a review. 

Building on these concepts, Shi et al.~(2022b) introduce a ``rank-based'' version 
 of dCov$({\bf X},{\bf Y})$ as the empirical counterpart of a particular case of the more general concept of   {\it generalized symmetric covariance},  a 
  direct definition of which is 
\begin{equation}\label{shietal1}
\tenq{\text{d}} \text{Cov} ({\bf X},{\bf Y})\coloneqq \text{dCov}({\bf F}_\pm^{\bf X}({\bf X}), {\bf F}_\pm^{\bf Y}({\bf Y})).
\end{equation}
    For $d_1=1=d_2$, ${\bf F}_\pm$ reduces to $2F -1$ and it is easy to see that 
 $$\tenq{\text{d}}\text{Cov}({X},{Y}) ={\text{d}}\text{Cov}(2F_X(X) -1, 2F_Y(Y)-1) = 4 {\text{d}}\text{Cov}(F_X(X),F_Y(Y))$$
 which, except for the inessential factor 4, is  the traditional population version of the rank-based  distance covariance.  
  Shi et al.~(2022b)  moreover allow for {\it scores}: denoting by  $J_1$ 
 and $J_2$ two   continuous mappings from~$[0,1)$ to ${\mathbb R}_+$, they define the $(J_1,J_2)$-{\it scored 
  distance covariance} as 
\begin{equation}\label{robshietal}
\tenq{\text{d}}\text{Cov}_{J_1,J_2}\left({\bf X},{\bf Y}\right)\coloneqq 
\text{dCov}
\left(
J_1\left(
\Vert {\bf F}^{\bf X}_\pm  ({\bf X})\Vert
\right)
\frac{{\bf F}^{\bf X}_\pm ({\bf X})}{\Vert {\bf F}^{\bf X}_\pm ({\bf X})\Vert}, 
 J_2\left(
 \Vert {\bf F}^{\bf Y}_\pm ({\bf Y})\Vert
 \right)
 \frac{{\bf F}^{\bf Y}_\pm ({\bf Y})}{\Vert {\bf F}^{\bf Y}_\pm ({\bf Y})\Vert}
 \right).
\end{equation}

In order to avoid unnecessarily heavy notation, and whenever the context   is clear, we use the same notation ${\bf F}^{(n)}_{\pm}$ for the marginal empirical transports  ${\bf F}^{(n)}_{{\bf X};\pm}$ and ${\bf F}^{(n)}_{{\bf Y};\pm}$ computed from the $\bf X$- and the~$\bf Y$-samples.  The empirical versions $\tenq{\text{d}} \text{Cov} ({\bf X}\n,{\bf Y}\n)$ and $\tenq{\text{d}}\text{Cov}_{J_1,J_2}\left({\bf X}\n,{\bf Y}\n\right)$ of~\eqref{shietal1} and~\eqref{robshietal}, based on a sample
  $({\bf X}^{(n)}_1,{\bf Y}^{(n)}_1)\ldots, ({\bf X}^{(n)}_n, {\bf Y}^{(n)}_n)
 $ 
 and involving the center-outward ranks and signs of  the $\bf X$- and the~$\bf Y$-samples, respectively, 
  are obtained by substituting in \eqref{shietal1} 
 the empirical distribution of the  $n$-tuple~$\big({\bf F}^{(n)}_\pm ({\bf X}_i\n),{\bf F}^{(n)}_\pm ({\bf Y}\n_i), i=1,\ldots,n\big)$ for the joint distributions of~$\big({\bf F}^{\bf X}_\pm ({\bf X}),{\bf F}^{\bf Y}_\pm ({\bf Y})\big)$ and, in \eqref{robshietal}, the empirical distribution of 
\begin{equation}\label{empiricalShi}
\left(
 J_1\left(
 \Vert {\bf F}\n_\pm  ({\bf X}\n_i)\Vert
 \right)
 \frac{{\bf F}\n_\pm ({\bf X}\n_i)}{\Vert {\bf F}\n_\pm ({\bf X}\n_i)\Vert}, 
 J_2\left(
 \Vert {\bf F}\n_\pm ({\bf Y}\n_i)\Vert
 \right)
 \frac{{\bf F}\n_\pm ({\bf Y}\n_i)}{\Vert {\bf F}\n_\pm ({\bf Y}\n_i)\Vert}
 , \ i=1,\ldots,n \right)
\end{equation}
 for the joint distribution of 
 $$
 J_1\left(
 \Vert {\bf F}^{\bf X}_\pm  ({\bf X})\Vert
 \right)
 \frac{{\bf F}^{\bf X}_\pm ({\bf X})}{\Vert {\bf F}^{\bf X}_\pm ({\bf X})\Vert}, 
 J_2\left(
 \Vert {\bf F}^{\bf Y}_\pm ({\bf Y})\Vert
 \right)
 \frac{{\bf F}^{\bf Y}_\pm ({\bf Y})}{\Vert {\bf F}^{\bf Y}_\pm ({\bf Y})\Vert}
 ,$$
  respectively.  
This yields, for scores 
$J_k(u)=\big({\bf F}^{-1}_{\chi^2_{d_k}} (u)
\big)^{1/2}$, $k=1,2$ ($F_{\chi^2_d} $ the $\chi^2_d$ distribution function with $d$ degrees of freedom) a van der Waerden version of~\eqref{shietal1}, for $J_1(u)=u=J_2(u)$,   the Wilcoxon distance covariance (also considered in Shi et al.~(2022a, 2022b), and, for~$J_1(u)=1=J_2(u)$,  a  sign-test-score version of the same.  

Substituting ${\bf F}^{(n)}_{0;\pm}$ for ${\bf F}^{(n)}_{\pm}$,    alternative versions of $\tenq{\text{d}} \text{Cov} ({\bf X}\n,{\bf Y}\n)$ and $\tenq{\text{d}}\text{Cov}_{J_1,J_2}\left({\bf X}\n,{\bf Y}\n\right)$  are obtained in an obvious way; they, however, do not fully deserve the qualification of ``rank-based''   distance covariance statistics.

\subsection{A bounded redescending influence function for rank-based distance covariance in ${\mathbb R}^d$\ ?}\label{sec22}

Neither the original distance covariance between random vectors nor its rank-based versions have been studied from the point of view of robustness. 
Extending the bivariate results developed by~[LRR] would be 
a natural continuation of their work. 
 In particular,  bounded redescending influence functions are likely to result from the following extensions of their   biloop strategy. 

The biloop transformation, actually, can be considered as a ``bivariate score function'' $\bf J$ acting on the modulus of the   center-outward distribution function, which is non-negative. Therefore, defining ${\bf J}_k$ as $\psi_\infty\circ J_k$, $k=1, 2$, with 
$${\psi_\infty^+}(u)\coloneqq\left(\!\! \begin{array}{c}c(1 + \cos (2\pi\tanh (u/c) +\pi))\\ \sin(2\pi \tanh(u/c)))
\end{array}
\!\!\right)$$
(${\psi_\infty^+}$'s  graph is the right-hand part of Figure~6), we suggest $\tenq{\text{d}}\text{Cov}_{{\bf J}_1,{\bf J}_2}({\bf X}\n,{\bf Y}\n)$ 
 (a  distance covariance between a $2d_1$- dimensional random vector 
 and a  $2d_2$-dimensional one) as a robustification of  $\tenq{\text{d}}\text{Cov}_{J_1,J_2}({\bf X}\n,{\bf Y}\n)$ (which is a  distance covariance between a $d_1$-    and a  $d_2$-dimensional random vector). Note that the {\it bi}loop nature of the transformation is taken care of by the fact that  the scored quantities ${\bf J}(\Vert {\bf F}_\pm \Vert)$, in~\eqref{robshietal}, are multiplied by the ``multivariate signs'' ${\bf F}_\pm / \Vert {\bf F}_\pm  \Vert$ which are uniform over the unit sphere. 
 
This yields, for  the same choices of $J_1$ and $J_2$ as above,  robustified versions of the van der Waerden, Wilcoxon, and sign-test-scores rank-based distance covariances. We conjecture 
 that the corresponding influence functions are bounded and redescending. 

\section{Distance covariance for directional data 
}\label{Sec3}

Section~\ref{SecRd} takes care of distance covariances for multivariate real data.  Further extensions, however, are highly desirable and have not been explored so far. In this section, we are sketching a possible extension to directional data, that is, to random directions distributed on hyperspheres.We conjecture that extensions to other compact non-linear Riemannian manifolds, such as tori and products of hyperspheres, are feasible using the same methodology, along with the nonparametric tools for Riemannian manifolds developed by \cite{HL24}. Here, we restrict the exposition to the case of  hyperspheres.  

\subsection{Data on  hyperspheres} 


Data with values on the sphere appear in a wide range of applications---environmental sciences (wind directions analysis in meterology), astronomy (directions of cosmic rays or stars), earth sciences (locations of an earthquake's epicentre on the surface of the earth), and biology (circadian rhythms, studies of animal navigation): see, e.g., the monograph by \cite{ley2017modern}. The specific nature and difficulty with such  data lies in the curvature of their sample space: hyperspheres or circles are non-linear manifolds. 

Although this type of data recently attracted a surge of activity,  the problem of testing independence between two directional variables seldom has been addressed, and some of the methods proposed, such as  in \cite{johnson2002tests},  for all their theoretical interest, are hardly implementable.   The ``robustness by data transformation paradigm'' emphasized in [LRR]  opens the door to  new developments and  novel independence testing procedures. In this section, we discuss some ideas  which, we believe, are paving the way to such developments.

Let us first introduce some  notation. Denote by~$\mathcal S^{d-1}$ the unit (hyper)sphere in \(\mathbb{R}^d\). The geodesic distance between  \(\mathbf{z}_1\) and~\( \mathbf{z}_2 \) in~\(\mathcal{S}^{d-1}\) is  \(d_\mathcal S (\mathbf{z}_1,\mathbf{z}_2) = |\mathrm{arccos}(\mathbf{z}_1^T\mathbf{z}_2)|\). When equipped with the geodesic distance, \(\mathcal{S}^{d-1}\) is a separable complete metric space, hence a Polish metric space with Borel \(\sigma\)-field \(\mathcal{B}^{d-1}\), say. The notation \(\mathrm{P}^\mathbf{Z}\) is used for the probability distribution of the $\mathcal{S}^{d-1}$-valued random direction~\(\mathbf{Z}\).  Finally, given a Riemannian manifold $\cal{M}$  
and a point $\mathbf{z} \in \mathcal{M}$ (this includes the hypersphere $\mathcal S^{d-1}$),   denote by~$T_{\mathbf{z}}{\mathcal{M}}$ the tangent space of \(\mathcal{M}\) at $\mathbf{z}$, and by $\partial {\cal M}$ the boundary of~${\cal M}$.

\subsection{Two problems} \label{Sec:DataSph}

Let \(\mathbf{X},\mathbf{Y}\) denote two random directions  defined on the same (unspecified) probability space, with values in $\mathcal{S}^{d_1-1}$ and~$\mathcal{S}^{d_2-1}$, respectively. 
 As in [LRR],  we are interested in testing the null hypothesis 
\begin{align}
    {\mathcal H}_0:~\mathbf{X}~\text{and}~\mathbf{Y}~\text{are independent,}
    \label{h0}
\end{align}
based on a sample   \((\mathbf{X}_1^{(n)},\mathbf{Y}_1^{(n)}),\dots,(\mathbf{X}_n^{(n)},\mathbf{Y}_n^{(n)})\) of~$n$ independent copies of~\((\mathbf{X},\mathbf{Y})\).
A natural question is: can we, as in [LRR],  use  distance covariance (preferably,  robust versions thereof) for the hypothesis testing problem~\eqref{h0}?    Natural as it is, this objective  nevertheless  runs into two serious  theoretical and methodo\-logical problems.
\begin{enumerate}
\item[(i)] {\it Problem 1 }(Hyperspheres are not  metric spaces of {\it strong negative type}) 
A   metric space \({\cal X} \) with metric~$d_{\cal X}$ is   of {\it negative type} if, for any \(n\ge2\), \({\bf x}_i\in {\cal X}\), and \(\alpha_i\in \mathbb{R}\), \(i = 1,2,\dots,n\) such that~\(\sum_{i=1}^n \alpha_i = 0\),  
 $\sum_{1\le i,j\le n}\alpha_i\alpha_j d_{\cal X}({\bf x}_i,{\bf x}_j) \le 0.$ 
 Let ${\cal P}_1({\cal X})$ denote the set of probability measures on $\cal X$ with finite  first-order moment. We say that~$({\cal X}, d_{\cal X})$ is of {\it strong negative type} if  (a)  $({\cal X}, d_{\cal X})$ is of negative type and (b) for any~${\rm P}_1, {\rm P}_2 \in {\cal P}_1({\cal X})$,
 $$\int_{{{\cal X}\times {\cal X}}}
  d_{\cal X}(\xb_1, \xb_2){\rm d} ({\rm P}_1 - {\rm P}_2)^{\otimes 2}(\xb_1, \xb_2) = 0$$ 
iff ${\rm P}_1 = {\rm P}_2$. \citet[Theorem 3.11 and Proposition 3.15]{lyons2013distance} shows that, for  two random vectors $\mathbf{X}$ and~$\mathbf{Y}$ with values in $({\cal X}, d_{\cal X})$ and $({\cal Y}, d_{\cal Y})$, respectively, and finite first-order moments, mutual independence  is equivalent to    \(\mathrm{dCov}(\mathbf{X},\mathbf{Y})= 0\) if and only if both  $\cal X$ and $\cal Y$ are metric spaces of  strong negative type. For the unit hypersphere ${\cal S}^{d-1}$ equipped with the geodesic distance, \cite{lyons2020strong} proves that a subset $\cal X$  of $\mathcal S^{d-1}$ is of strong negative type if and only if it contains at most one pair of antipodal points. As a consequence,    distance covariance tests of independence between  directional variables $\Xb$ and $\Yb$ are inappropriate unless 
the support of $\Xb$ and the support of~$\Yb$ both contain at most one pair of antipodal points. This, at first sight, precludes the application of distance-covariance-based tests on (hyper)spheres.


\item[(ii)] {\it Problem 2 }(Distribution-freeness)   Assuming that  point (i) can be taken care of,  a typical  issue, if a test based on distance covariance is to be implemented,  
is that  the (asymptotic) null distribution of~$\mathrm{dCov}^2$ depends on the (typically unspecified) distributions of \(\mathbf{X}\) and \(\mathbf{Y}\). This dependence is  usually hard to deal with (both theoretically and numerically): see Remark~4.4 in  \cite{DebSen23} for a  discussion of this issue. If a distribution-free version of the distance covariance statistic can be defined for $d\geq 2$ (similar to  the definition for $d=1$ in  Section~\ref{sec21}), then critical values can be obtained, with arbitrary precision,  by simulations. 
\end{enumerate} 

In the next two sections, we show how [LRR]'s robustness by transformation principle, combined with the theory of optimal transportation on  hyperspheres (more generally, on compact Riemannian manifolds) and the introduction of directional ranks and signs,  provides a solution to these two problems.

\subsection{Measure transportation and directional distance co\-variance}\label{sphrank}

Let us start with a brief definition of the concepts to be used.   The theory of measure transportation,\footnote{Note that the expression {\it optimal} (for squared Euclidean distance transportation costs) {\it transportation} (OT) only makes sense in the context of probability measures with finite second order moments---a condition which is auto\-matically satisfied for variables with values on compact manifolds. }   which can be traced back to \cite{monge1781memoire} and   \cite{kantorovich1960mathematical},  has been successfully applied in a variety of areas, including statistics, economics, machine learning, computer science, to mention only a few: see,  e.g., \citep{G17,hallin2022measure,la2023some,MLLVL25} and references therein. 

In statistics, based on measure transportation,  \cite{Chernoetal17} (under the name of {\it Monge-Kantorovich ranks}) and \cite{hallin2021center}  (under the name of {\it center-outward ranks})   introduce  novel notions of ranks (and signs and quantiles)  in \(\mathbb{R}^d\) that enjoy all the  properties expected from such concepts---among which (for the ranks and the signs) distribution-freeness. The approach and the terminology we are adopting here is that of \cite{hallin2021center, hallin2024nonparametric}.  

These measure-transportation-based techniques have been used 
  to construct new distribution-free  tests of independence in \textit{Euclidean spaces}, defining, in particular, rank-based distance covariance test statistics for vector independence: see Section~\ref{sec21}. Based on optimal transportation (since variances in the context are finite), \cite{hallin2024nonparametric}  on hyperspheres  and  \cite{HL24} on more general compact Riemannian manifolds also define distribution-free ranks and signs.  
 The directional distance covariance and the  testing procedures we are proposing below are building on these theoretical and methodological developements. 

\subsection{Population concepts: directional distribution and quantile functions} \label{sec:pop}

Before proceeding further, let us summarize some key results of~\cite{hallin2024nonparametric}, who 
propose directional concepts of distribution and quantile functions inducing a distribution-specific system of curvilinear parallels and (hyper)meridians on the sphere. 
For any distributions P and~Q on~$( \mathcal{S}^{d-1},\mathcal{B}^{d-1})$, let \(\mathcal{S}(\mathrm{P},\mathrm{Q})\)  denote the set of all measurable   maps \(\mathbf{S}: \mathcal{S}^{d-1}\rightarrow \mathcal{S}^{d-1}\) such that for all~$V\in \mathcal{B}^{d-1}$, \((\mathbf{S}\#\mathrm{P})(V)\coloneqq \mathrm{P}(\mathbf{S}^{-1}(V))=\mathrm{Q}(V)\)---in the measure transportation terminology, the set of transport maps $\bf S$ \textit{pushing} P \textit{forward to} Q.  The optimal transportation (OT) problem  on~\(\mathcal{S}^{d-1}\)   consists in minimizing, for given P and Q and  over ${\bf S}\in\mathcal{S}(\mathrm{P},\mathrm{Q})$, the quantity 
\begin{equation} C({\bf S})\coloneqq\int_{\mathcal{S}^{d-1}\times\mathcal{S}^{d-1}}c(\mathbf{z},{\bf S}(\mathbf{z}))\mathrm{dP}(\mathbf{z}) \quad\text{where}\quad 
c(\mathbf{z}_1, \mathbf{z}_2)\coloneqq d_\mathcal S^2(\mathbf{z}_1,\mathbf{z}_2)/2. \label{cost} 
\end{equation} 
Proposition 1 in \cite{hallin2024nonparametric}  establishes  the existence and ${\rm P}$-a.s. uniqueness of the solution~${\bf S}_{\rm P, Q}$ and Proposition~2 shows that it is a homeomorphism (hence, is continuously invertible) between $\mathcal{S}^{d-1}$ and $\mathcal{S}^{d-1}$;   call {\it geodesic Wasserstein distance} between P and Q the value $C({\bf S}_{\rm P, Q})$ of the minimum.  
When~Q is the uniform \({\rm P}^{\bf U}\)  over~\(\mathcal{S}^{d-1}\), call ${\bf F}={\bf F}_{\rm P}$ and ${\bf Q} = {\bf Q}_{\rm P}\coloneqq {\bf F}_{\rm P}^{-1}$  the {\it directional distribution function} and the  {\it directional quantile function of}~P, respectively (parallel to~${\bf F}_{\rm P}$ and ${\bf Q}_{\rm P}$, the notation ${\bf F}_{\bf Z}$ and ${\bf Q}_{\bf Z}$ will be used in an obvious way for ${\bf Z}\sim {\rm P}$).
%
%
%
%
These definitions are a natural extension of the classical univariate ones, where the distribution function~F of a real-valued random variable $Z\sim {\rm P}^Z$ is pushing ${\rm P}^Z$ forward to the uniform  $\mathrm{U}_{[0, 1]}$ over the unit interval. 

\cite{hallin2024nonparametric} first define collections of nested regions of $\mathcal{S}^{d-1}$ with \({\rm P}^{\bf U}\)-probability contents~\(\tau\in [0, 1]\) centered at some submanifold $\mathcal{S}^{d-1}_0$ with zero ${\rm P}^{\bf U}$-probability content.  These regions have the interpretation of quantile regions of order $\tau\in [0,1]$ for the uniform ${\rm P}^{\bf U}$, with median set~$\mathcal{S}^{d-1}_0$. For the sake of simplicity, we follow \cite{hallin2024nonparametric} and restrict  $\mathcal{S}^{d-1}_0$ to be a singleton~$\{{\boldsymbol\theta}\}$;  other choices are possible, though (e.g., the equator associated with some pole~${\boldsymbol\theta}$---see \cite{HL24}). Since ${\rm P}^{\bf U}$ is fully symmetric, the choice of ${\boldsymbol\theta}$  is to be based on~$\rm P$, and \cite{hallin2024nonparametric} propose for ${\boldsymbol\theta}$ the image by $\bf F$ of $\rm P$'s Fr\' echet mean 
  ${\boldsymbol\theta}^{\rm P}_{\scriptscriptstyle{\text{\rm Fr\' echet}}}\!\!\coloneqq  \argmin_{{\bf z}\in\mathcal{S}^{d-1}} \int_{\mathcal{S}^{d-1}}c({\bf z}, x)\mathrm{dP} (x)$ 
%
 (in case of multiple Fr\' echet means, pick an arbitrary one or choose some centroid of the Fr\' echet mean set).  
%
%
%
%
%
 Due to the rotational symmetry of \({\rm P}^{\bf U}\) with respect to \(\mathbf{F}({{\boldsymbol\theta}})\), the collection of nested spherical caps centered at~\(\mathbf{F}({{\boldsymbol\theta}})\) with~\({\rm P}^{\bf U}\)-probability contents \(\tau\) constitutes a natural family of  quantile regions for the uniform  
 \({\rm P}^{\bf U}\). 
\cite{hallin2024nonparametric} then define the $\tau$-quantile region~$ \mathbb{C}_{\tau}=  \mathbb{C}_{\tau}^{\rm P}$ of~\(\mathrm{P}\) as the  image by $\bf Q$ of the  $\tau$-quantile region  $\mathbb{C}^{\rm{U}}_{\tau}$  of ${\rm P}^{\bf U}$. More precisely, 
\begin{align*}
    \mathbb{C}^{\mathrm{U}}_{\tau}&\coloneqq \{\mathbf{u}\in \mathcal{S}^{d-1}:F_*(\mathbf{u}^T\mathbf{F}({\boldsymbol\theta})) \ge 1 - \tau\},\\
     \mathbb{C}_{\tau}=\mathbb{C}_{\tau}^{\rm P}&\coloneqq \mathbf{Q}(\mathbb{C}^{\mathrm{U}}_\tau) = \{\mathbf{z}\in \mathcal{S}^{d-1}:F_*(\mathbf{F}(\mathbf{z})^T\mathbf{F}({\boldsymbol\theta})) \ge 1 - \tau\},
\end{align*}
with boundaries (quantile {\it contours}) 
\begin{align*}
    \mathcal{C}^\mathrm{U}_\tau&\coloneqq \{\mathbf{u}\in \mathcal{S}^{d-1}:F_*(\mathbf{u}^T\mathbf{F}({\boldsymbol\theta})) = 1 - \tau\},\\
  \mathcal{C}_\tau^{\rm P}=  \mathcal{C}_\tau&\coloneqq \mathbf{Q}(\mathcal{C}^\mathrm{U}_\tau)=\{\mathbf{z}\in \mathcal{S}^{d-1}:F_*(\mathbf{F}(\mathbf{z})^T\mathbf{F}({\boldsymbol\theta})) = 1 - \tau\},
\end{align*}
where \(F_*\) is the distribution function of $\mathbf{U}^T \mathbf{F}({\boldsymbol\theta})$ where $\mathbf{U}\sim {\rm P}^{\bf U}$, which takes the form
\begin{equation*}
    F_*(u)\coloneqq \int_{-1}^u(1-s^2)^{(d-3)/2}\mathrm{d}s / \int_{-1}^1(1-s^2)^{(d-3)/2}\mathrm{d}s,~-1\le u \le 1
 \end{equation*}
 and adjusts the size of the spherical caps  $\mathbb{C}^{\mathrm{U}}_{\tau}$ to achieve probability content $\tau$.

\subsection{Empirical counterparts: directional ranks, signs, and quantiles} \label{EmpRS}

Let \(\mathbf{Z}^{(n)}=\{{\bf Z}^{(n)}_1,\dots,{\bf Z}^{(n)}_n\}\) denote a sample of $n$ i.i.d.\ copies of  ${\bf Z}\sim \mathrm{P}$.  As in ${\mathbb R}^d$, two empirical versions of ${\bf F}\equiv {\bf F}_{\rm P}$,  ${\bf F}\n_{0}$ and ${\bf F}\n$,  can be defined, with the difference that ${\bf F}\n$ here  is the result of a two-step procedure of which ${\bf F}\n_{0}$ is the first step. Both ${\bf F}\n_{0}$ and ${\bf F}\n$ are Glivenko-Cantelli consistent, but only ${\bf F}\n$ yields quantile contours, ranks, and signs. Let us describe the two-step construction of~${\bf F}\n$.


The first step 
 involves the estimation  of a pole ${\boldsymbol\theta}$ for $\rm P$---for instance, the empirical Fr\' echet\linebreak mean~ ${\boldsymbol\theta}^{(n)} = {\boldsymbol\theta}^{(n)}_{\scriptscriptstyle{\text{\rm Fr\' echet}}}$---along with an estimation  ${\bf F}^{(n)}_0$ of $\bf F$. The latter is obtained as the mapping from the sample~\(\mathbf{Z}^{(n)}\) to an adequate grid \(\mathfrak{G}^{(n)}_0=\{{\scriptstyle{\mathfrak{G}}}^{(n)}_{0;1},\dots, {\scriptstyle{\mathfrak{G}}}^{(n)}_{0;n}\}\subset \mathcal{S}^{d-1}\) minimizing, over the set~$\Pi^{(n)}$ of all permutations~$\pi$ of the integers $\{1,\ldots,n\}$, 
$\sum_{i=1}^{n}c({\bf Z}_i^{(n)}, {\scriptstyle{\mathfrak{G}}}_{0;\pi(i)}^{(n)})$ (with $c(\cdot,\cdot)$ defined in \eqref{cost}). The solution ${\bf F}\n_{0}$ of this minimization problem is the    optimal transport pushing the empirical distribution of the sample forward to the empirical distribution of the grid $\mathfrak{G}^{(n)}_0$. The only condition required 
 for the consistency of  ${\bf F}\n_{0}$ is  the weak convergence, as~$n\to\infty$, of the   empirical distribution over $\mathfrak{G}^{(n)}_0$  to the uniform over~$\mathcal{S}^{d-1}$; to fix the ideas, let us assume that $\mathfrak{G}^{(n)}_0$ is obtained as an i.i.d.\  $n$-tuple of uniform  over~${\cal S}^{d-1}$ variables.  This minimization  is an optimal pairing problem for which efficient algorithms are available.  
The resulting~${\bf F}_0^{(n)}$ does not allow for the construction of empirical quantile regions and contours, ranks, or signs. However,   it determines a data-driven pole $\thetat \coloneqq {\bf F}^{(n)}_0({\boldsymbol\theta}^{(n)})$ for the uniform, around which the grid to be used in the second  step  will be  based, and provides each observation ${\bf Z}_i^{(n)}$ with a {\it latitude}~$1-\langle {\bf F}_0^{(n)}({\bf Z}_i^{(n)}), \thetat 
\rangle$ and  a {\it (hyper)longitude}---the intersection of the geodesic connecting $ \thetat $ and ${\bf F}_0^{(n)}({\bf Z}_i^{(n)})$ with the equatorial region associated with  $\thetat $, i.e., the $(d-2)$-hypersphere ${\cal S}^{d-2}$ of the equatorial hyperplane determined by~$\thetat$;  this equatorial hyperplane coincides with the translation $T_{\thetat}{\cal S}^{d-1}-\thetat$ of the tangent space at $\thetat$.

 The second step involves a second grid, $\mathfrak{G}^{(n)}_{\thetat}$, based, as in ${\mathbb R}^d$,  a factorization of $n$ into $n_Rn_S~\!+~\! n_0$ with $n_0 < \min (n_R, n_S)$. This second grid consists of the $n_Rn_S$    intersections of $n_S$ uniformly distributed (hyper)meridians (geodesics connecting $\thetat$ and $-\thetat$)  with a uniform array of $n_R$ {\it parallels} (a parallel is the collection of points with given latitude), along with $n_0$ copies of the pole~$\thetat $. 
%
%
Denote by ${\bf F}^{(n)}\equiv {\bf F}^{(n)}_{\bf Z}$ the optimal transport from the empirical distribution of the sample to the empirical distribution over this second grid. The directional ranks $R_i\n$ (taking values~$1,\ldots,n_R$)   and   signs~${\bf S}_i\n$ (taking values in the $(d-2)$-hypershere of the equatorial hyperplane determined by~$\thetat$) to be used below can be defined as functions of ${\bf F}\n_{\Zb}({\bf Z}_i\n)$. We refer to Section~4.2 of \cite{hallin2024nonparametric} for explicit formulas.

\subsection{A rank-based directional distance covariance  test of independence}\label{indep-test}

Let us show how the ranks and signs obtained by letting ${\bf Z}\n_i = {\bf X}\n_i $ and ${\bf Z}\n_i = {\bf Y}\n_i $ in the procedures described in~Section~\ref{EmpRS} allow for constructing  distance covariance statistics  that  simultaneously  take care of the two problems---distribution-freeness and the fact that the hypersphere  is not of strong  negative type---mentioned in Section~\ref{Sec:DataSph}. 


Recall  that  \(\mathbf{F}_{\bf X}\) and  \(\mathbf{F}_{\bf Y}\) are homeomorphic transformations of \(\mathcal{S}^{d_1-1}\) and  \(\mathcal{S}^{d_2-1}\), respectively. Therefore, 
testing the mutual independence of   \({\bf X}\) and \({\bf Y}\) is equivalent to testing the mutual independence of the transformed variables~\(\mathbf{F}_{\bf X}({\bf X})\) and \(\mathbf{F}_{\bf Y}({\bf Y})\). This suggests  considering a distribution-free test based on the empirical distance covariance between the transformed observations  
\begin{equation}\label{transfdata0}
\mathbf{F}^{(n)}_{0;{\bf X}}({\bf X}^{(n)}_1),\ldots ,  \mathbf{F}^{(n)}_{0;{\bf X}}({\bf X}^{(n)}_n)
\quad\text{ and}\quad 
\mathbf{F}^{(n)}_{0;{\bf Y}}({\bf Y}^{(n)}_1),\ldots ,  \mathbf{F}^{(n)}_{0;{\bf Y}}({\bf Y}^{(n)}_n)
\end{equation}
\setcounter{equation}{3}
or
\begin{equation}\label{transfdata}
\mathbf{F}^{(n)}_{\bf X}({\bf X}^{(n)}_1),\ldots ,  \mathbf{F}^{(n)}_{\bf X}({\bf X}^{(n)}_n)
\quad\text{ and}\quad 
\mathbf{F}^{(n)}_{\bf Y}({\bf Y}^{(n)}_1),\ldots ,  \mathbf{F}^{(n)}_{\bf Y}({\bf Y}^{(n)}_n)
\end{equation}
where $\mathbf{F}^{(n)}_{0;{\bf X}}$ and $\mathbf{F}^{(n)}_{0;{\bf Y}}$ stand for the optimal transports obtained, for ${\bf X}^{(n)}_1\!,\ldots, {\bf X}^{(n)}_n$and~${\bf Y}^{(n)}_1\!,\ldots, {\bf Y}^{(n)}_n\!$, respectively,  in Step 1 of Section~\ref{EmpRS}, $\mathbf{F}^{(n)}_{\bf X}$ and $\mathbf{F}^{(n)}_{\bf Y}$ for the optimal transports obtained in 
Step 2 (with, using obvious notation for the poles $\widehat{\boldsymbol\theta}\n_{{\bf X}}$ 
and $\widehat{\boldsymbol\theta}\n_{{\bf Y}}$,  grids $\mathfrak{G}^{(n)}_{\widehat{\boldsymbol\theta}\n_{{\bf X}}}$ over  \(\mathcal{S}^{d_1-1}\) for ${\bf X}\n$ and $\mathfrak{G}^{(n)}_{\widehat{\boldsymbol\theta}\n_{{\bf Y}}}$ over  \(\mathcal{S}^{d_2-1}\)  for ${\bf Y}\n$). Note that these $n$-tuples of transformed data are uniformly distributed  over the $n!$ permutations of their respective grids, hence distribution-free; contrary to $\mathbf{F}^{(n)}_{\bf X}$ and $\mathbf{F}^{(n)}_{\bf Y}$, however,   $\mathbf{F}^{(n)}_{0;X}$ and $\mathbf{F}^{(n)}_{0;Y}$ do not define ranks and signs in~Section~\ref{EmpRS}.

Now,   the transformed data \eqref{transfdata0} and~\eqref{transfdata} still take values in 
  $\mathcal{S}^{d_1-1}$ and $\mathcal{S}^{d_2-1}$, which contain infinitely many  pairs of antipodal points, thus failing to be of the strongly negative type. Compu\-ting distance covariances between them, therefore, remains theoretically meaningless. To circumvent this problem,
 we propose a further  transformation or, more precisely, a class of further bijective transformations of~\(\mathbf{F}^{(n)}_{\bf X}({\bf X}_i^{(n)})\) and~\(\mathbf{F}^{(n)}_{\bf Y}({\bf Y}_i^{(n)})\) (of~\(\mathbf{F}^{(n)}_{0;{\bf X}}({\bf X}_i^{(n)})\) and \(\mathbf{F}^{(n)}_{0;{\bf Y}}({\bf Y}_i^{(n)})\)), $i=1,\ldots, n$ mapping  \(\mathcal{S}^{d_1-1}\) and  \(\mathcal{S}^{d_2-1}\) to 
  Euclidean spaces; being bijective, these transformations  are preserving the null hypothesis of independence, while allowing for a meaningful notion of distance covariance. 


For any point ${\bf z}\in \mathcal{S}^{d-1}$, there exist  homeomorphisms $\varphi: \mathcal{S}^{d-1}\setminus\{-{\bf z}\}\rightarrow {K}^{d-1}$, where ${K}^{d-1}$ is some open subset of the tangent space $T_{\bf z}{\mathcal{S}^{d-1}}$ of $\mathcal{S}^{d-1}$ at ${\bf z}$.  
Letting  $\varphi_1$ and $\varphi_2$ denote such  transformations for ${\bf z}_1\in  \mathcal{S}^{d_1-1}$ and ${\bf z}_2\in  \mathcal{S}^{d_2-1}$, respectively consider 
\begin{equation} \varphi_1\circ \mathbf{F}\n_{0;{\bf X}}\quad\text{and} \quad \varphi_2\circ \mathbf{F}\n_{0;{\bf Y}}
\quad\text{ or }\quad 
\varphi_1\circ \mathbf{F}\n_{{\bf X}}\quad\text{and} \quad \varphi_2\circ \mathbf{F}\n_{{\bf Y}}\label{Eq.varphi}.
 \end{equation}
By construction, $\varphi _1$ and  $\varphi _2$   map the   observation $({\bf X}\n_i\!, {\bf Y}\n_i)\!\in\!  \big(\mathcal{S}^{d_1-1}\!\setminus\!\{-{\bf z}_1\}\big)\!\times\! \big( \mathcal{S}^{d_2-1}\!\setminus\!\{-{\bf z}_2\}\big)$\linebreak   to~${K}^{d_1-1}\times {K}^{d_2-1} \subset T_{{\bf z}_1}{\mathcal{S}^{d_1-1}}\times T_{{\bf z}_2}{\mathcal{S}^{d_2-1}}$, which is a product of Euclidean spaces    
 over which  $\mathrm{dCov}$ is defined as usual.  Natural choices for ${\bf z}_1$ and  ${\bf z}_2$ are  the empirical poles ${{\thetab}\n_{\bf X}}$ and ${{\thetab}\n_{\bf Y}}$ (values of ${\bf z}_1$ and  ${\bf z}_2$ coinciding with any $-{\bf X}\n_i$ and $-{\bf Y}\n_i$ are to be avoided lest the corresponding observation is lost). We then propose the following definition of a directional distance covariance statistic. 

\begin{definition}[Directional distance covariance]
  (i) Define the $(\varphi_1,\varphi_2)$-directional distance co- \begin{enumerate}
  \item[]variance of~$({\bf X}, {\bf Y})$ with values in $\mathcal{S}^{d_1-1}\! \times \mathcal{S}^{d_2-1}$ as  
    \begin{equation}
  \tenq{\rm d}\text{Cov}_{\varphi_1, \varphi_2}({\bf X}, {\bf Y}) \coloneqq \mathrm{dCov}\left(\varphi_1(\mathbf{F}_{\bf X}(\mathbf{X})), \varphi_2(\mathbf{F}_{\bf Y}(\mathbf{Y}))\right), \label{Eq.Ddcov}
\end{equation}
where $\varphi_1$ and $\varphi_2$  are bijective mappings from $\mathcal{S}^{d_1-1}\!\,\setminus\!\{-{\boldsymbol\theta}_{\bf X}\}$ and $\mathcal{S}^{d_1-2}\!\,\setminus\!\{-{\boldsymbol\theta}_{\bf Y}\}$ to  open subsets~$K^{d_1-1}$ of $T_{{\boldsymbol\theta}_{\bf X}}{\mathcal{S}^{d_1-1}}$ and~$K^{d_2-1}$ of $T_{{\boldsymbol\theta}_{\bf Y}}{\mathcal{S}^{d_2-1}}$, respectively. 
\item[(ii)]With  bijective mappings $\varphi_1$ and $\varphi_2$ from 
$\mathcal{S}^{d_1-1}\!\,\setminus\!\{-\widehat{\boldsymbol\theta}\n_{\bf X}\}$ and $\mathcal{S}^{d_2-1}\!\,\setminus\!\{-\widehat{\boldsymbol\theta}\n_{\bf Y}\}$ 
to  open subsets~$K^{d_1-1}$ of $T_{\widehat{\boldsymbol\theta}\n_{\bf X}}{\mathcal S}^{d_1-1}$ and~$K^{d_2-1}$ of $T_{\widehat{\boldsymbol\theta}\n_{\bf Y}}{\mathcal{S}}^{d_2-1}$, 
empirical versions of   \eqref{Eq.Ddcov} 
 are the  statistics~$   \tenq{\rm d}\text{Cov}_{\varphi_1, \varphi_2}({\bf X}\n, {\bf Y}\n) \coloneqq \mathrm{dCov}\left(\varphi_1(\mathbf{F}\n_{\bf X}(\mathbf{X}\n)), \varphi_2(\mathbf{F}\n_{\bf Y}(\mathbf{Y}\n))\right)$ (which is rank-based),  
 and  
 $  \tenq{\rm d}\text{Cov}_{0;\varphi_1, \varphi_2}({\bf X}\n ,{\bf Y}\n) \coloneqq \mathrm{dCov}\left(\varphi_1(\mathbf{F}\n_{0;\bf X}(\mathbf{X}\n)), \varphi_2(\mathbf{F}\n_{0;\bf Y}(\mathbf{Y}\n))\right)
$; both are distribution-free under the null hypothesis ${\mathcal H}_0$ of independence between $\bf X$ and $\bf Y$.
\end{enumerate}\end{definition}

Distribution-free tests of independence then can be constructed, which are rejecting ${\mathcal H}_0$ for ``large values'' of test statistics of the form
\begin{equation}\label{tildeW}
\tenq{W}\n _{\varphi_1,\varphi_2}
\coloneqq n\, \tenq{\rm d}\text{Cov}^2_{\varphi_1, \varphi_2}({\bf X}\n, {\bf Y}\n)
\quad\text{or}\quad
\tenq{W}\n_{0; \varphi_1,\varphi_2}
\coloneqq 
n\, \tenq{\rm d}\text{Cov}^2_{0;\varphi_1, \varphi_2}({\bf X}\n ,{\bf Y}\n) 
\end{equation}
 (the factor $n$ entails  convergence, as $n\to\infty$ of the null distributions of $\tenq{W}\n $ and  $\tenq{W}_0\n$  to a nondegenerate limit, as in Theorem 3.1 of Shi et al. (2020)).




Both $\tenq{W}\n _{\varphi_1,\varphi_2}$ and $\tenq{W}\n_{0; \varphi_1,\varphi_2}$ have their own merits. On the one hand, $\tenq{W}\n_{0; \varphi_1,\varphi_2}$ is of less computational complexity than $\tenq{W}\n _{\varphi_1,\varphi_2}$ since it  does not require the second optimal transport in Step 2 of Section~\ref{EmpRS}. On the other hand, unlike $\tenq{W}\n _{\varphi_1,\varphi_2}$,  it does not fully qualify as ``rank-based.''    As for the choice of~$\varphi_{1}$ and~$\varphi_{2}$,  one may  privilege bounded functions  yielding robustness features, pretty much  in the same spirit as the function $\psi_\infty$ introduced in [LRR]. 

A natural specification  of  $\varphi_{1}$ and $\varphi_2$ in $\tenq{W}\n _{\varphi_1,\varphi_2}$, yielding a class of tests that are closely related to the  classical examples of rank- and sign-based scores in the univariate setting  and share the same spirit as the measure-transportation-based inference procedures (estimation and testing) developed in   \cite{ hallin2022center,HLVL23,hallin2023center} for ${\mathbb R}^d$-valued  data is \begin{align}
	\varphi_1\left(\mathbf{F}_{\Xb}^{(n)}(\mathbf{X}^{(n)}_i)\right)= \varphi_{1 1}\left(\frac{R\n_{\mathbf{X}, i}}{n_R + 1}\right) \varphi_{1 2}(\mathbf{S}\n_{\mathbf{X}, i}) \quad \text{and} \quad 	\varphi_2\left(\mathbf{F}_{\Yb}^{(n)}(\mathbf{Y}^{(n)}_i)\right)=& \varphi_{2 1}\left(\frac{R\n_{\mathbf{Y}, i}}{n_R + 1}\right) \varphi_{2 2}(\mathbf{S}\n_{\mathbf{Y}, i}),\nonumber\\ 
	&\qquad i=1,2,\dots,n
	\label{ourphi}
\end{align}
(with obvious notation for~$R\n_{\mathbf{X}, i}$, $\mathbf{S}\n_{\mathbf{X}, i}$,  $R\n_{\mathbf{Y}, i}$, and~$\mathbf{S}\n_{\mathbf{Y},i}$)
 where  we assume that  
 $\varphi_{1 1}$ and $\varphi_{2 1}$ are continuous and bijective functions mapping~[0,~\!1)  to $\mathbb{R}$ while~$\varphi_{1 2}$ and $\varphi_{2 2}$ are continuous and bijective  mappings from $ \mathcal{S}^{d_1-2}$ and $ \mathcal{S}^{d_2-2}$ to the tangent spaces $T_{\widehat{\boldsymbol\theta}\n_{\Xb}}{\mathcal{S}^{d_1-1}}$ and $T_{\widehat{\boldsymbol\theta}\n_{\Yb}}{\mathcal{S}^{d_2-1}}$, respectively. 

%

Classical choices for the score functions $\varphi$ in~\eqref{ourphi} are 
\begin{enumerate}
\item[--]  (Wilcoxon scores) $\varphi_{\ell 1}(u) = u$ and $\varphi_{\ell 2}(\ub) = \ub$ for~$\ell =1,2$;
\item[--]  (Gaussian or van der Waerden scores) $\varphi_{\ell 1}(u) = \big[F^{-1}_{\chi^2_{d}} (u)\big]^{1/2}$ and  $\varphi_{\ell 2}(\ub) = \ub$ for~$\ell =1,2$ where $F^{-1}_{\chi^2_{d}}$ stands for the quantile function of a chi-square distribution with $d$ (either $d_1$ or~$d_2$) degrees of freedom.
\end{enumerate}
Both choices yield  homeomorphisms from  $\mathcal{S}^{d-1}\!\setminus\!\{-\widehat{\boldsymbol\theta}\n_{\Xb}\}$ and $\mathcal{S}^{d-1}\!\setminus\!\{-\widehat{\boldsymbol\theta}\n_{\Yb}\}$  to the open unit disks~${\mathbb D}^{d-1}_{\widehat{\boldsymbol\theta}\n_{\Xb}}$ of $T_{\widehat{\boldsymbol\theta}\n_{\Xb}}{\mathcal{S}^{d-1}}$ and~${\mathbb D}^{d-1}_{\widehat{\boldsymbol\theta}\n_{\Yb}}$ of $T_{\widehat{\boldsymbol\theta}\n_{\Yb}}{\mathcal{S}^{d-1}}$, respectively. Both are fine since we are now in a Euclidean space rather than in $\mathcal{S}^{d-1}$, which takes care of \textit{Problem 1} of  Section~\ref{Sec:DataSph}.


As for  \textit{Problem 2},    $\tenq{W}\n _{\varphi_1,\varphi_2}$ and $\tenq{W}\n_{0; \varphi_1,\varphi_2}$ (as defined in \eqref{tildeW})  by construction are 
 jointly distribution-free under the null hypothesis of independence.  
 
 Sign-test scores  $\varphi_{\ell 1}(u) = 1$ and $\varphi_{\ell 2}(\ub) = \ub$ for~$\ell =1,2$ are  another popular choice. However, they are an example of a non-bijective  transformation of the original observations, yielding  a test (based on the test statistic $\tenq{W}\n_{\varphi_1,\varphi_2}$)  that  remains distribution-free but fails to detect some types of dependencies.

\section{Conclusion}

Building on the measure-transportation-based notions of directional ranks and signs defined in \cite{hallin2024nonparametric}, we show how the ``robustness via transformation'' principle emphasized by [LRR] extends beyond the case of bivariate independence and also applies in higher-dimension Euclidean spaces and on compact manifolds. The case of directional variables (taking values on (hyper)spheres) is given special attention:  although (hyper)spheres are not  metric spaces of the strong negative type, we 
  introduce a class of pertinent directional distance covariance test statistics and  illustrate their applicability.  We hope that this note will open the door to further research directions in line with [LRR]. 
%
%

\bibliographystyle{chicago}
\bibliography{reference.bib}


\end{document}